\documentclass[a4paper,11pt]{article}

\usepackage{authblk}
\usepackage{verbatim}

\usepackage{graphicx}% Include figure files
\usepackage{dcolumn}% Align table columns on decimal point
\usepackage{amstext,amsmath,amssymb,amsfonts}
\usepackage{bm}% bold math

\newcommand{\captionfonts}{\small}
\makeatletter  % Allow the use of @ in command names
\long\def\@makecaption#1#2{%
  \vskip\abovecaptionskip
  \sbox\@tempboxa{{\captionfonts #1: #2}}%
  \ifdim \wd\@tempboxa >\hsize
    {\captionfonts #1: #2\par}
  \else
    \hbox to\hsize{\hfil\box\@tempboxa\hfil}%
  \fi
  \vskip\belowcaptionskip}
\makeatother   % Cancel the effect of \makeatletter

\usepackage{color}
\definecolor{lightblue}{rgb}{0.2,0.2,0.7}
\definecolor{darkblue}{rgb}{0,0.25,0.5}
\definecolor{redbrown}{rgb}{0.875,0.25,0.125}
\definecolor{darkgreen}{rgb}{0,0.5,0}

\newcommand{\bra}[1]{\ensuremath{\langle #1 \vert}}
\newcommand{\ket}[1]{\ensuremath{\vert #1  \rangle}}
\newcommand{\braket}[2]{\ensuremath{\langle  #1 \vert #2  \rangle}}
\renewcommand{\b}[1]{\ensuremath{\mathbf{#1}}}

\newcommand{\T}{\ensuremath{\text{T}}}

\newcommand{\E}{\ensuremath{\text{E}}}
\newcommand{\V}{\ensuremath{\text{V}}}
\newcommand{\Cov}{\ensuremath{\text{Cov}}}
\newcommand{\FN}{\ensuremath{\text{FN}}}

\renewcommand{\L}{\ensuremath{\text{L}}}

\renewcommand{\d}{\ensuremath{\text{d}}}
\newcommand{\s}{\ensuremath{\text{s}}}
\renewcommand{\c}{\ensuremath{\text{c}}}
\newcommand{\block}{\ensuremath{\text{b}}}

\newcommand{\f}{\ensuremath{\text{f}}}
\renewcommand{\i}{\ensuremath{\text{i}}}
\renewcommand{\L}{\ensuremath{\text{L}}}

\usepackage{fancyhdr}

\makeatletter
\renewcommand\paragraph{\@startsection{paragraph}{4}{\z@}%
  {-3.25ex\@plus -1ex \@minus -.2ex}%
  {1.5ex \@plus .2ex}%
  {\normalfont\normalsize\bfseries}}
\makeatother

\begin{document}

\title{Introduction to the variational and diffusion Monte Carlo methods}

%%%%%%%%%%%%%%%%%%%%%%%%%%%%%%%%%%%%%%%%%%%%%%%%%%%%%%%%%%%%%%%%%%%%%%%%%%%%%
\author{
Julien Toulouse$^{1,2}$, Roland Assaraf$^{1,2}$, C. J. Umrigar$^3$\\
$^1$Sorbonne Universit\'es, UPMC Univ Paris 06, UMR 7616, Laboratoire de Chimie Th\'eorique, F-75005 Paris, France\\
$^2$CNRS, UMR 7616, Laboratoire de Chimie Th\'eorique, F-75005 Paris, France\\
$^3$Laboratory of Atomic and Solid State Physics, Cornell University, Ithaca, New York 14853, USA\\
}

%%%%%%%%%%%%%%%%%%%%%%%%%%%%%%%%%%%%%%%%%%%%%%%%%%%%%%%%%%%%%%%%%%%%%%%%%%%%%
\date{August 10, 2015}

\maketitle
\tableofcontents

\newpage
\section*{Abstract}
\addcontentsline{toc}{section}{Abstract}
We provide a pedagogical introduction to the two main variants of real-space quantum Monte Carlo methods for electronic-structure calculations: variational Monte Carlo (VMC) and diffusion Monte Carlo (DMC). Assuming no prior knowledge on the subject, we review in depth the Metropolis-Hastings algorithm used in VMC for sampling the square of an approximate wave function, discussing details important for applications to electronic systems. We also review in detail the more sophisticated DMC algorithm within the fixed-node approximation, introduced to avoid the infamous Fermionic sign problem, which allows one to sample a more accurate approximation to the ground-state wave function. Throughout this review, we discuss the statistical methods used for evaluating expectation values and statistical uncertainties. In particular, we show how to estimate nonlinear functions of expectation values and their statistical uncertainties.

\vspace{1cm}
\noindent
{\bf Keywords:} quantum Monte Carlo, electronic-structure calculations, Metropolis-Hastings algorithm, fixed-node approximation, statistical methods.

\newpage
This chapter provides a pedagogical introduction to the two main variants of real-space quantum Monte Carlo (QMC) methods for electronic-structure calculations: variational Monte Carlo (VMC) and diffusion Monte Carlo (DMC). For more details of these methods, see, e.g., Refs.~\cite{HamLesRey-BOOK-94,NigUmr-INC-98,Umr-INC-99,FouMitNeeRaj-RMP-01,ReyCepAldLes-JCP-82,UmrNigRun-JCP-93}. For reviews on applications of QMC methods in chemistry and condensed-matter physics, see, e.g., Refs.~\cite{AusZubLes-CR-12,KolMit-RPP-11}.

\section{Variational Monte Carlo}

\subsection{Basic idea}

The idea of the VMC method~\cite{Mil-PR-65,CepCheKal-PRB-77} is simply to calculate the multidimensional integrals appearing in quantum mechanics using a Monte Carlo numerical integration technique\footnote{To the best of our knowledge, the first calculation of multidimensional integrals appearing in quantum mechanics by using Monte Carlo methods was done by Conroy~\cite{Con-JCP-64}.}. The quantity of greatest interest is the variational energy associated with a Hamiltonian $\hat{H}$ and a wave function $\Psi$, which can be written as
\begin{equation}
E_\text{v} = \frac{\bra{\Psi} \hat{H} \ket{\Psi}}{\braket{\Psi}{\Psi}} = \frac{\int \d\b{R} \, \Psi(\b{R})^2 E_\L(\b{R})}{\int \d\b{R} \, \Psi(\b{R})^2} = \int \d\b{R} \, \rho(\b{R}) E_\L(\b{R}),
\end{equation}
where $E_\L(\b{R}) = (H\Psi(\b{R}))/\Psi(\b{R})$ is the \textit{local energy} depending on the $3N$ coordinates $\b{R}$ of the $N$ electrons, and $\rho(\b{R})= \Psi(\b{R})^2/\int \d\b{R} \Psi(\b{R})^2$ is the normalized probability density.
For simplicity of notation we have assumed that $\Psi(\b{R})$ is real valued; the extension to complex $\Psi(\b{R})$ is straightforward.
The variational energy can be estimated as the average value of $E_\L(\b{R})$ on a sample of $M$ points $\b{R}_k$ sampled from the probability density $\rho(\b{R})$,
\begin{equation}
E_\text{v} \approx \overline{E}_\L = \frac{1}{M} \sum_{k=1}^{M} E_\L(\b{R}_k).
\label{Elav}
\end{equation}
In practice, the points $\b{R}_k$ are sampled using the Metropolis-Hastings algorithm~\cite{MetRosRosTelTel-JCP-53,Has-B-70}.

The advantage of this approach is that it does not use an analytical integration involving the wave function, and thus does not impose severe constraints on the form of the wave function. The wave functions usually used in QMC are of the Jastrow-Slater form,
\begin{equation}
\Psi(\b{R}) = J(\b{R}) \Phi(\b{R}),
\end{equation}
where $J(\b{R})$ is a Jastrow factor and $\Phi(\b{R})$ is a Slater determinant or a linear combination of Slater determinants\footnote{In QMC, it is convenient to use wave functions in which the values of the spin coordinates have been fixed, so $\Psi$ is a function of the spatial coordinates $\b{R}$ only.}. The Jastrow factor is generally of the form $J(\b{R})=e^{f(\b{R})}$.  It depends explicitly on the interparticle distances $r_{ij}$, allowing for an efficient description of the so-called electron ``dynamic'' correlation.

In practice, the VMC method has two types of errors:
\begin{itemize}
\item a {\it systematic error}, due to the use of an approximate wave function (as in other wave-function methods),
\item a {\it statistical uncertainty}, due to the sampling of finite size $M$ (which is specific to Monte Carlo methods).
\end{itemize}
Of course, the variational energy is an upper bound of the exact ground-state energy, but the systematic error is generally unknown, since its determination requires knowing the exact solution. By contrast, the statistical uncertainty can be easily estimated by the usual statistical techniques. For this, let us examine more closely the meaning of Eq.~(\ref{Elav}). The average of the local energy  $\overline{E}_\L$ on a finite sample is itself a random variable, taking different values on different samples. The central limit theorem establishes that, if $E_\L(\b{R}_k)$ are random variables that are {\it independent} (i.e. not correlated) and {\it identically distributed}, with finite expected value $\E[E_\L]$ and finite variance, $\V[E_\L]=\E[(E_\L-E_\text{v})^2]$, then in the large $M$ limit  the probability distribution of the random variable $\overline E_\L$ converges (in the mathematical sense of convergence in distribution) to a Gaussian (or normal) distribution of expected value $\E[E_\L]$  and variance $\V[E_\L]/M$,
\begin{subequations}
\begin{equation}
\E\left[\,\overline{E}_\L\right] = \E[E_\L] = E_\text{v},
\end{equation}
\begin{equation}
\V\left[\overline{E}_\L\right] = \frac{\V[E_\L]}{M}.
\end{equation}
\end{subequations}
This means that $\overline{E}_\L$ is an {\it estimator} of $E_\text{v}$ with a statistical uncertainty which can be defined by the standard deviation of its Gaussian distribution
\begin{equation}
\sigma\left[\overline{E}_\L\right] = \sqrt{\V\left[\overline{E}_\L\right]}=\sqrt{\frac{\V[E_\L]}{M}}.
\end{equation}
The meaning of this standard deviation is that the desired expected value $E_\text{v}$ has a probability of 68.3\% of being in the interval $\left[\overline{E}_\L     - \sigma, \overline{E}_\L + \sigma\right]$, a probability of 95.5\% of being in the interval $\left[\overline{   E}_\L - 2\sigma, \overline{E}_\L  + 2\sigma\right]$, and a probability of 99.7\% of being in the interval $\left[\overline{E}_\L- 3\sigma, \overline{E}_\L + 3\sigma\right]$. Note that, if the variance $\V[E_\L]$ is infinite but the expected value $\E[E_\L]$ is finite, then the law of large numbers guarantees the convergence of $\overline{E}_\L$ to $\E[E_\L]$ when $M \to \infty$ but with a statistical uncertainty which is more difficult to estimate and which decreases more slowly than $1/\sqrt{M}$.

It is important to note that the statistical uncertainty decreases as $1/\sqrt{M}$ {\it independently of the dimension of the problem}. This is in contrast to deterministic numerical integration methods for which the convergence of the integration error deteriorates with the spatial dimension $d$. For example, Simpson's integration rule converges as $1/M^{(4/d)}$ (provided the integrand has up to $4^{th}$-order derivatives), so that for $d>8$ Monte Carlo methods are more efficient for large $M$.

The statistical uncertainty is reduced if the variance of the local energy $\V[E_\L]$ is small. In the limit that $\Psi$ is an exact eigenfunction of $\hat{H}$, the local energy $E_\L$ becomes exact, independent of $\b{R}$, and thus its variance $\V[E_\L]$ and the statistical uncertainty of $\overline{E}_\L$ vanish. This is known as the \textit{zero-variance} property. Since the systematic error (or bias) of the variational energy  $\Delta E = E_\text{v} - E_0$ (where $E_0$ is the exact energy) also vanishes in this limit, there is a zero-bias property as well.  For these reasons, a great deal of effort has been expended on developing robust and efficient wave-function optimization methods.

\subsection{Estimation of the statistical uncertainty}

In practice, the probability density $\rho(\b{R})$ is sampled with the Metropolis-Hastings algorithm which provides a sequence of points $\b{R}_k$ correctly distributed according to $\rho(\b{R})$ but {\it sequentially (or serially) correlated} (i.e. non independent). This is a consequence of each point being sampled from a probability distribution conditional on the previous point.  One can define an {\it autocorrelation time} (defined more precisely later) that is roughly speaking the average time for points to decorrelate. This sequential correlation must be taken into account when using the central limit theorem for evaluating the statistical uncertainty. This is done using the {\it blocking} technique, which is described next.

Let us consider a sequence of $M$ realizations $X_k$ (sequentially correlated) of a random variable $X$ of expected value $\E[X]$ and of variance $\V[X]$. For example, $X$ could be the local energy $E_\L$. We divide this sequence into $M_\block$ successive blocks of $M_\s$ steps each. The {\it block average} $\overline{X}_\block$ is
\begin{equation}
\overline{X}_\block =  \frac{1}{M_\s} \sum_{k=1}^{M_\s} X_{k}.
%\overline{X}_b =  \frac{1}{M_\s} \sum_{k=1}^{M_\s} X_{k+ (b-1) M_\block}.
\label{Xblock}
\end{equation}
The expected value of $\overline{X}_\block$ is also the expected value of $X$, i.e. $E\left[\overline{X}_\block\right]=E[X]$, but its variance is not simply $V[X]/M_\s$ since the variables $X_k$ are not independent. We can now define the {\it global average} $\overline{X}$ of the whole sample as the average over all the blocks of the block averages
\begin{equation}
\overline{X} = \frac{1}{M_\block} \sum_{b=1}^{M_\block} \overline{X}_b,
\end{equation}
where $\overline{X}_b$ with a math subscript ``$b$" indicates the block average for the $b^{th}$ block (whereas $\overline{X}_\block$ with a Roman subscript ``b" indicates the generic random variable). The global average $\overline{X}$ is another random variable with the same expected value as $X$, i.e. $\E\left[\overline{X}\right]=\E\left[\overline{X}_\block\right]=\E[X]$. If the length of the blocks is large compared to the autocorrelation time then the block averages $\overline{X}_b$ can be considered as being independent, and the variance of the global average is simply
\begin{equation}
\V \left[ \overline{X} \right] = \frac{\V\left[\overline{X}_\block\right]}{M_\block},
\end{equation}
which leads to the statistical uncertainty of $\overline{X}$
\begin{equation}
\sigma\left[\,\overline{X}\right]=\sqrt{\V\left[\overline{X}\right]} = \sqrt{\frac{\V\left[{\overline{X}_\block}\right]}{M_\block}}.
\end{equation}
In practice, the statistical uncertainty on a finite sample is calculated as
\begin{equation}
\sigma\left[\overline{X}\right] \approx \sqrt{\frac{1}{M_\block-1} \left( \frac{1}{M_\block} \sum_{b=1}^{M_\block} {\overline{X}_b}^2 - \left( \frac{1}{M_\block} \sum_{b=1}^{M_\block} \overline{X}_b\right)^2 \right)},
\end{equation}
where the $M_\block-1$ term appearing instead of $M_\block$ is necessary to have an unbiased estimator of the standard deviation on the sample (see the appendix).
It takes into account the fact that the computed variance is relative to the sample average rather than the true expected value.

Finally, let us examine the variance $\V\left[\overline{X}_\block\right]$. Since the variables $X_k$ are not independent, the expansion of $\V\left[\overline{X}_\block\right]$ involves the covariances between the variables
\begin{equation}
\V\left[\overline{X}_\block\right] = \frac{1}{M_\s^2} \sum_{k,l} \Cov[X_k,X_l] = \frac{\V[X]}{M_\s} + \frac{2}{M_\s^2} \sum_{k<l} \Cov[X_k,X_l] = T_\c \frac{\V[X]}{M_\s},
\end{equation}
defining the autocorrelation time of $X$
\begin{equation}
T_\c = 1 + \frac{2}{\V[X]M_\s} \sum_{k<l} \Cov[X_k,X_l].
\end{equation}
The autocorrelation time is equal to $1$ in the absence of correlation between the variables, i.e. $\Cov[X_k,X_l]=0$ for $k\not=l$, but can be large in the presence of sequential correlation. It is instructive to express the statistical uncertainty as a function of $T_\c$
\begin{equation}
\sigma\left[\overline{X}\right]=\sqrt{T_\c \frac{\V[X]}{M_\s M_\block}}  = \sqrt{T_\c \frac{\V[X]}{M}},
\label{sigmaavecTc}
\end{equation}
where $M=M_\s M_\block$ is the total size of the sample. The expression~(\ref{sigmaavecTc}) allows one to interpret $T_\c$ as a factor giving the number of effectively independent points in the sample, $M_\text{eff} = M/T_\c$. In practice, it is useful to calculate the autocorrelation time as $T_\c =M_\s \V\left[\overline{X}_\block\right]/\V[X]$ and check whether the length of the blocks is large enough for a correct estimation of the statistical uncertainty, e.g. $M_\s > 100 \, T_\c$. If $M_\s$ is not much greater than $T_\c$, then the statistical uncertainty $\sigma\left[\overline{X}\right]$ and the autocorrelation time $T_\c$ will be underestimated.

In the appendix, we further explain how to estimate the statistical uncertainty of nonlinear functions of expectation values, which often occur in practice.

\subsection{Calculation cost}

The calculation cost required to reach a given statistical uncertainty $\sigma\left[\overline{X}\right]$ is
\begin{equation}
t = t_\s M = t_\s \frac{T_\c \V[X]}{\sigma\left[\overline{X}\right]^2}
\end{equation}
where $t_\s$ is the calculation time per iteration. The $1/\sigma\left[\overline{X}\right]^2$ dependence implies that decreasing the statistical uncertainty by a factor of $10$ requires to increase the computational time by a factor of $100$. This quadratic dependence directly stems from the central limit theorem and seems unavoidable\footnote{Quasi Monte Carlo methods~\cite{Caf-ActNum-98} can in some cases achieve a convergence rate of ${\cal O} (\ln(M)/M)$ rather than ${\cal O} (1/\sqrt{M})$.  However, they have not been used for QMC applications, in part because in QMC the sampled distributions, for systems with more than a few electrons, are very highly peaked.}. However, one can play with the three other parameters:
\begin{itemize}
\item $T_\c$ depends on the sampling algorithm and on the random variable $X$. For efficient algorithms such as Umrigar's one~\cite{Umr-PRL-93,Umr-INC-99}, the autocorrelation time of the local energy is close to $1$ and little further improvement seems possible;
\item $t_\s$ is usually dominated by the cost of evaluating $X$. For the local energy, the evaluation cost depends on the form of the wave function;
\item $\V[X]$ depends on the choice of the random variable $X$ with its associated probability distribution, the only constraint being that the expected value $\E[X]$ must equal the expectation value of the observable (otherwise, this is a biased estimator). The choice of a good probability distribution is usually called {\it importance sampling}.
Even for a fixed probability distribution, it is possible to use various estimators for $X$, some of which have smaller variance than others, since one has the freedom to add any quantity with zero expectation value.
This has been exploited to construct improved estimators for diverse observables~\cite{AssCaf-PRL-99,AssCaf-JCP-00,AssCaf-JCP-03,AssCafSce-PRE-07,TouAssUmr-JCP-07}. There is often a compromise to be found between a low computation time per iteration $t_\s$ and a low variance $\V[X]$.
\end{itemize}

\subsection{Sampling technique}

The probability density, $\rho(\b{R})= \Psi(\b{R})^2/\int \d\b{R} \Psi(\b{R})^2$, is generally complicated and cannot be sampled by direct methods such as the transformation method or the rejection method.  Instead, the Metropolis-Hastings (or generalized Metropolis) algorithm, which can be used to sample any known probability density, is used.  It employs a stochastic process, more specifically, a Markov chain.

\paragraph{Stochastic process}
A {\it stochastic process} represents the evolution -- say in ``time'' -- of a random variable. It is described by a trajectory of successive points $\b{R}_1$, $\b{R}_2$, ..., $\b{R}_M$ with an associated probability distribution $P(\b{R}_M,...,\b{R}_2,\b{R}_1)$. The idea of evolution in time can be made more explicit by decomposing the probability of the whole trajectory in products of the conditional probability of having a particular point knowing that all the previous points have already been realized. For example, for $M=3$, the probability of the trajectory is
\begin{equation}
P(\b{R}_3,\b{R}_2,\b{R}_1) = P(\b{R}_3|\b{R}_2,\b{R}_1) P(\b{R}_2|\b{R}_1) P(\b{R}_1).
\end{equation}

\paragraph{Markov chain}
A {\it Markov chain} is a stochastic process for which the conditional probability for the transition to a new point $\b{R}_k$ depends only on the previous point $\b{R}_{k-1}$
\begin{equation}
P(\b{R}_k|\b{R}_{k-1},...,\b{R}_1) = P(\b{R}_k|\b{R}_{k-1}),
\end{equation}
i.e. the process ``forgets'' the way it arrived at point $\b{R}_{k-1}$. The probability of a trajectory can thus be simply written as, e.g. for $M=3$,
\begin{equation}
P(\b{R}_3,\b{R}_2,\b{R}_1) = P(\b{R}_3|\b{R}_2) P(\b{R}_2|\b{R}_1) P(\b{R}_1),
\end{equation}
and $P(\b{R}_\f|\b{R}_\i)$ is called the transition probability from point $\b{R}_\i$ to point $\b{R}_\f$. Note that, in general, the transition probability can depend on time (measured by the index $k$). We will consider here only the case of a stationary Markov chain for which the transition probability is time independent.

In the following, we will use notation corresponding to the case of states $\b{R}_k$ in a continuous space (``integrals'' instead of ``sums''), but we will ignore the possibly subtle mathematical differences between the continuous and discrete cases, and we will often use the vocabulary of the discrete case (e.g., ``matrix''). The transition probability matrix, $P$ is a {\it stochastic matrix}, i.e., it has the following two properties:
\begin{subequations}
\begin{equation}
P(\b{R}_\f|\b{R}_\i) \geq 0 \;\;\;\; (\text{non negativity}),
\label{nonnegativite}
\end{equation}
\begin{equation}
\int \d\b{R}_\f  \, P(\b{R}_\f|\b{R}_\i) =1 \;\;\;\; (\text{column normalization}).
\label{normalisation}
\end{equation}
\end{subequations}
The second property expresses the fact that the probability that a point $\b{R}_\i$ is somewhere at the next step must be $1$.  The eigenvalues of a stochastic matrix are between $0$ and $1$, and there is at least one eigenvalue equal to $1$.
The latter property is a consequence of the fact that, for a column-normalized matrix, the vector with all components equal to one is a left eigenvector with eigenvalue $1$.
The target probability distribution $\rho(\b{R})$ is sampled by constructing a Markov chain converging to $\rho(\b{R})$. A necessary condition is that the distribution $\rho(\b{R})$ is a (right) eigenvector of $P(\b{R}_\f|\b{R}_\i)$ with the eigenvalue $1$
\begin{equation}
\int \d\b{R}_\i  \, P(\b{R}_\f|\b{R}_\i) \rho(\b{R}_\i) = \rho(\b{R}_\f) = \int \d\b{R}_\i  \, P(\b{R}_\i|\b{R}_\f) \rho(\b{R}_\f)  \;\;\;\;\;\; \forall \, \b{R}_\f,
\label{balance}
\end{equation}
where the second equality simply comes from the normalization condition~(\ref{normalisation}). Eq.~(\ref{balance}) is a {\it stationarity condition} for $\rho(\b{R})$. It means that if we start from the target distribution $\rho(\b{R})$ then we will continue to sample the same distribution by applying the Markov chain. However, we need more than that. We want that any initial distribution $\rho_{\text{ini}}(\b{R})$, e.g., a  delta function at some initial point, evolves to the target stationary distribution $\rho(\b{R})$ by repeated applications of the transition matrix
\begin{eqnarray}
\lefteqn{\lim_{M\to\infty}\int \d\b{R}_1  \, P^{M}(\b{R}|\b{R}_1) \rho_{\text{ini}}(\b{R}_1) =} &&
\nonumber\\
&& \lim_{M\to\infty}\int \d\b{R}_1\d\b{R}_2...\d\b{R}_{M}  \, P(\b{R}|\b{R}_{M}) P(\b{R}_M|\b{R}_{M-1})...P(\b{R}_2|\b{R}_{1}) \rho_{\text{ini}}(\b{R}_1) = \rho(\b{R}),
\end{eqnarray}
i.e. $\rho(\b{R})$ must be the dominant eigenvector of $P$ (the unique eigenvector of largest eigenvalue). If the stationarity condition~(\ref{balance}) is satisfied then this will always be the case except if $P$ has several eigenvectors with eigenvalue $1$. One can show that the matrix $P$ has only one eigenvector of eigenvalue $1$ if $P$ is a primitive matrix, i.e. if there is an integer $n\geq1$ such that all the elements of the matrix $P^n$ are strictly positive, $P^n(\b{R}_k|\b{R}_l) >0, \;\;\forall \; \b{R}_k,\b{R}_l$.
This means that it must be possible to move between any pair of states $\b{R}_k$ and $\b{R}_l$ in $n$ steps.
This ensures that all states can be visited, and that the Markov chain converges to the unique stationary distribution $\rho(\b{R})$. The Markov chain is then said to be {\it ergodic}.

In practice, instead of imposing the stationarity condition~(\ref{balance}), the Markov matrix is constructed by imposing the more stringent {\it detailed balance} condition,
\begin{equation}
P(\b{R}_\f|\b{R}_\i) \rho(\b{R}_\i) = P(\b{R}_\i|\b{R}_\f) \rho(\b{R}_\f),
\label{detailedbalance}
\end{equation}
which forces the probability flux between the two states $\b{R}_\i$ and $\b{R}_\f$ to be the same in both directions. This is a sufficient (but not necessary) condition for $\rho(\b{R})$ to be the stationary distribution. A Markov chain satisfying condition~(\ref{detailedbalance}) is said to be reversible.

In practice, a Markov chain is realized by a {\it random walk}. Starting from an initial point $\b{R}_1$ (or walker) -- i.e. a delta-function distribution $\delta(\b{R}-\b{R}_1)$ -- sample the second point $\b{R}_2$ by drawing from the probability distribution $P(\b{R}_2|\b{R}_{1})$, then a third point $\b{R}_3$ by drawing from $P(\b{R}_3|\b{R}_{2})$, and so on. After disregarding a certain number of iterations $M_\text{eq}$ corresponding to a transient phase called {\it equilibration}, the random walk samples the stationary distribution $\rho(\b{R})$ in the sense that $\rho(\b{R}) = \E[\delta(\b{R}-\b{R}_k)] \approx (1/M) \sum_{k=1}^M \delta(\b{R}-\b{R}_k)$ and the averages of the estimators of the observables of interest are calculated. The rate of convergence to the stationary distribution $\rho(\b{R})$ and the autocorrelation times of the observables are determined by the second largest eigenvalue of the matrix $P$ (see, e.g., Ref.~\cite{GilRicSpi-BOOK-96}). The random walk must be sufficiently long so as to obtain a representative sample of the states making a non negligible contribution to the expected values. If the transitions between states belonging to two contributing regions of the space of states are too improbable, as may happen for example for dissociated atoms, then there is a risk that the random walk remains stuck in a region of space and seems converged, even though the true stationary distribution is not yet reached.
To avoid this problem, smart choices for the transition matrix can be crucial in various applications of Monte Carlo methods~\cite{Wol-PRL-89,MelSan-PRE-05}.

\paragraph{Metropolis-Hastings algorithm}

In the Metropolis-Hastings algorithm~\cite{MetRosRosTelTel-JCP-53,Has-B-70}, one realizes a Markov chain with the following random walk. Starting from a point $\b{R}_\i$, a new point $\b{R}_\f$ is determined in two steps:
\begin{enumerate}
\item a temporary point $\b{R}_\f^\prime$ is proposed with the probability $P_{\text{prop}}(\b{R}_\f^\prime|\b{R}_{\i})$,
\item the point $\b{R}_\f^\prime$ is accepted (i.e. $\b{R}_\f = \b{R}_\f^\prime$) with probability $P_{\text{acc}}(\b{R}_\f^\prime|\b{R}_{\i})$, or rejected (i.e. $\b{R}_\f = \b{R}_\i$) with probability $P_{\text{rej}}(\b{R}_\f^\prime|\b{R}_{\i}) = 1 - P_{\text{acc}}(\b{R}_\f^\prime|\b{R}_{\i})$
\end{enumerate}
The corresponding transition probability can be written as
\begin{equation}
P(\b{R}_\f|\b{R}_{\i}) =
  \begin{cases}
   P_{\text{acc}}(\b{R}_\f|\b{R}_{\i}) P_{\text{prop}}(\b{R}_\f|\b{R}_{\i}) & \text{if } \b{R}_\f \neq \b{R}_{\i} \\
   1 - \int \d\b{R}_\f^\prime \; P_{\text{acc}}(\b{R}_\f^\prime|\b{R}_{\i}) P_{\text{prop}}(\b{R}_\f^\prime|\b{R}_{\i})        & \text{if } \b{R}_\f = \b{R}_{\i}
  \end{cases}
\end{equation}
or, in a single expression,
\begin{equation}
P(\b{R}_\f|\b{R}_{\i}) = P_{\text{acc}}(\b{R}_\f|\b{R}_{\i}) P_{\text{prop}}(\b{R}_\f|\b{R}_{\i}) + \left[  1 - \int \d\b{R}_\f^\prime \; P_{\text{acc}}(\b{R}_\f^\prime|\b{R}_{\i}) P_{\text{prop}}(\b{R}_\f^\prime|\b{R}_{\i}) \right] \delta(\b{R}_{\i}-\b{R}_\f).
\label{Ptransition}
\end{equation}
The proposal probability $P_{\text{prop}}(\b{R}_\f|\b{R}_{\i})$ is a stochastic matrix, i.e. $P_{\text{prop}}(\b{R}_\f|\b{R}_{\i})\geq 0$ and\\ $\int \d\b{R}_\f P_{\text{prop}}(\b{R}_\f|\b{R}_{\i})=1$, ensuring that $P(\b{R}_\f|\b{R}_{\i})$ fulfils the non-negativity condition~(\ref{nonnegativite}). The second term in Eq.~(\ref{Ptransition}) with the delta function ensures that $P(\b{R}_\f|\b{R}_{\i})$ fulfils the normalization condition~(\ref{normalisation}). The acceptance probability is chosen so as to fulfil the detailed balance condition~(\ref{detailedbalance}), for $\b{R}_\f \neq \b{R}_{\i}$,
\begin{equation}
\frac{P_{\text{acc}}(\b{R}_\f|\b{R}_{\i})}{P_{\text{acc}}(\b{R}_\i|\b{R}_{\f})} = \frac{P_{\text{prop}}(\b{R}_\i|\b{R}_{\f}) \rho(\b{R}_{\f})}{P_{\text{prop}}(\b{R}_\f|\b{R}_{\i}) \rho(\b{R}_{\i})}.
\end{equation}
Several choices are possibles. The choice of Metropolis {\it et al.}~\cite{MetRosRosTelTel-JCP-53} maximizes the acceptance probability
\begin{equation}
P_{\text{acc}}(\b{R}_\f|\b{R}_{\i}) = \min \left\{ 1, \frac{P_{\text{prop}}(\b{R}_\i|\b{R}_{\f}) \rho(\b{R}_{\f})}{P_{\text{prop}}(\b{R}_\f|\b{R}_{\i}) \rho(\b{R}_{\i})} \right\}.
\label{PaccMetropolis}
\end{equation}
The acceptance probability is not a stochastic matrix, even though both the proposal and the total Markov matrices are stochastic.
Since only the ratio $\rho(\b{R}_{\f})/\rho(\b{R}_{\i})$ is involved in Eq.~(\ref{PaccMetropolis}), it is not necessary to calculate the normalization constant of the probability density $\rho(\b{R})$.
It is clear that the acceptance probability of Eq.~(\ref{PaccMetropolis}) is optimal, but there is considerable scope for ingenuity in choosing a proposal probability $P_{\text{prop}}(\b{R}_\f|\b{R}_{\i})$ that leads to a small autocorrelation time.

\paragraph{Choice of the proposal probability}

The original paper of Metropolis {\it et al.}~\cite{MetRosRosTelTel-JCP-53} employed a symmetric proposal matrix, in which case the proposal matrix drops out of the formula for the acceptance.
The advantage of having a nonsymmetric proposal matrix was pointed out by Hastings~\cite{Has-B-70}.
One has a lot of freedom in the choice of the proposal probability $P_{\text{prop}}(\b{R}_\f|\b{R}_{\i})$. The only constraints are that $P_{\text{prop}}(\b{R}_\f|\b{R}_{\i})$ must be a stochastic matrix leading to an ergodic Markov chain and that it must be possible to efficiently sample $P_{\text{prop}}(\b{R}_\f|\b{R}_{\i})$ with a direct sampling method. The proposal probability determines the average size of the proposed moves $\b{R}_{\i} \to \b{R}_\f$ and the average acceptance rate of these moves. In order to reduce sequential correlation, one has to make moves as large as possible but with a high acceptance rate. In practice, for a given form of the proposal matrix, there is a compromise to be found between the average size of the proposed moves and the average acceptance rate.

The simplest choice for $P_{\text{prop}}(\b{R}_\f|\b{R}_{\i})$ is a distribution that is uniform inside a small cube $\Omega(\b{R}_{\i})$ centered in $\b{R}_{\i}$ and of side length $\Delta$ and zero outside
\begin{equation}
P_{\text{prop}}(\b{R}_\f|\b{R}_{\i}) =
  \begin{cases}
   \frac{1}{\Delta^{3N}}  \;\; \text{if } \b{R}_\f \in \Omega(\b{R}_{\i}) \\
   0                      \;\; \text{elsewhere }.
  \end{cases}
\label{Ppropcube}
\end{equation}
In practice, a move according to Eq.~(\ref{Ppropcube}) is proposed,
\begin{equation}
\b{R}_\f = \b{R}_\i +\frac{\Delta}{2} \; \bm{\chi},
\end{equation}
where $\bm{\chi}$ is a vector of $3N$ random numbers drawn from the uniform distribution between $-1$ and $1$. The size of the cube $\Delta$ can be adjusted so as to minimize the autocorrelation time of the local energy, but the latter remains large and the sampling is inefficient.

Clever choices use information from the distribution $\rho(\b{R})$, in particular its local gradient, to guide the sampling. A choice for $P_{\text{prop}}(\b{R}_\f|\b{R}_{\i})$ which would lead to large moves with an acceptance probability equal to $1$ would be $P_{\text{prop}}(\b{R}_\f|\b{R}_{\i})=\rho(\b{R}_\f)$, independently from $\b{R}_{\i}$, but we would then be back to the initial problem of sampling a complicated distribution $\rho(\b{R})$. A good choice for $P_{\text{prop}}(\b{R}_\f|\b{R}_{\i})$ is the Green function of the Fokker-Planck equation in the short-time approximation
\begin{equation}
P_{\text{prop}}(\b{R}_\f|\b{R}_{\i}) = \frac{1}{\left(2\pi \tau\right)^{3N/2}} e^{- \frac{\left( \b{R}_\f - \b{R}_\i -\b{v}(\b{R}_\i) \tau\right)^2}{2\tau}},
\label{PpropFokker-Planck}
\end{equation}
where $\b{v}(\b{R}) = \bm{\nabla} \Psi(\b{R}) / \Psi(\b{R})$ is called the {\it drift velocity} of the wave function and $\tau$ is the time step which can be adjusted so as to minimize the autocorrelation time of the local energy. In practice, a move according to Eq.~(\ref{PpropFokker-Planck}) is proposed
\begin{equation}
\b{R}_\f = \b{R}_\i +\b{v}(\b{R}_\i) \tau + \bm{\eta},
\label{Ri_to_Rf}
\end{equation}
where $\bm{\eta}$ is a vector of $3N$ random numbers drawn from the Gaussian distribution of average $0$ and standard deviation $\sqrt{\tau}$. The term $\bm{\eta}$ describes an isotropic Gaussian diffusion process (or Wiener process). The term $\b{v}(\b{R}_\i) \tau$ is a drift term which moves the random walk in the direction of increasing $|\Psi(\b{R})|$.

The optimal size of the move is smaller in regions where $\b{v}(\b{R})$ is changing rapidly. For example, $\b{v}(\b{R})$ has a discontinuity at the nuclear positions. Hence, it is more efficient to make smaller moves for electrons in the core than for electrons in the valence regions.
In doing this, care must be taken to ensure the detailed balance condition.
An elegant solution is provided in the VMC algorithm of Refs.~\cite{Umr-PRL-93,Umr-INC-99} where the electron moves are made in spherical coordinates centered on the nearest nucleus and the size of radial moves is proportional to the distance to the nearest nucleus.
In addition, the size of the angular moves gets larger as one approaches a nucleus.  This algorithm allows one to achieve, in many cases, an autocorrelation time of the local energy close to $1$.

\paragraph{Expectation values}
The expectation value of an operator $\hat{O}$ can be computed by averaging the corresponding local value $O(\b{R}_\f) = \bra{\b{R}_\f} \hat{O} \ket{\Psi}/\Psi(\b{R}_\f)$ at the Monte Carlo points $\b{R}_\f$ after the accept/reject step. A somewhat smaller statistical error can be achieved by instead averaging
\begin{eqnarray}
P_{\text{acc}}(\b{R}_\f|\b{R}_{\i}) \; O(\b{R}_\f) + (1-P_{\text{acc}}(\b{R}_\f|\b{R}_{\i})) \; O(\b{R}_\i),
\end{eqnarray}
regardless of whether the proposed move is accepted or rejected.

\paragraph{Moving the electrons all at once or one by one?}
So far we have assumed that, for a many-electron system, all the electrons are moved and then this move is accepted or rejected in a single step. In fact, it is also possible to move the electrons one by one, i.e. move the first electron, accept or reject this move, then move the second electron, accept or reject this move, and so on. In this case, the transition probability for $N$ electrons can be formally decomposed as
\begin{eqnarray}
P(\b{R}_\f|\b{R}_{\i}) &=& P(\b{r}_{1,\f}\b{r}_{2,\f}...\b{r}_{N,\f}|\b{r}_{1,\f}\b{r}_{2,\f}...\b{r}_{N,\i}) \times
\nonumber\\
&&... \times P(\b{r}_{1,\f}\b{r}_{2,\f}...\b{r}_{N,\i}|\b{r}_{1,\f}\b{r}_{2,\i}...\b{r}_{N,\i}) \times P(\b{r}_{1,\f}\b{r}_{2,\i}...\b{r}_{N,\i}|\b{r}_{1,\i}\b{r}_{2,\i}...\b{r}_{N,\i}),
\label{Ponebyone}
\end{eqnarray}
where each one-electron transition probability (knowing that the other electrons are fixed) is made of a proposal probability and an acceptance probability just as before. If each one-electron transition probability satisfies the stationary  condition~(\ref{balance}), then the global transition probability satisfies it as well.

Moving the $N$ electrons one by one requires more calculation time than moving the electrons all at once, since the wave function must be recalculated after each move to calculate the acceptance probability. The calculation time does not increase by a factor of $N$ as one may naively think but typically by a factor of 2 if the value of the wave function is recalculated in a clever way after an one-electron move. For example, for Slater determinants, one can use the matrix determinant lemma in conjunction with the Sherman-Morrison formula (see, e.g., Ref.~\cite{PreTeuVetFla-BOOK-92}) to efficiently recalculate the values of the determinants when only one row or column has been changed. In spite of the increase in the calculation time, it has been repeatedly shown in the literature (see, e.g., Refs.~\cite{CepCheKal-PRB-77,Umr-PRL-93,LopMaDruTowNee-PRE-06,LeeConNemLopDru-PRE-11}) that, for systems with many electrons, moving the electrons one by one leads to a more efficient algorithm: larger moves can be made for the same average acceptance, so the points $\b{R}_k$ are less sequentially correlated and the autocorrelation time of the local energy is smaller (by a factor larger than the one necessary for compensating the increase of the calculation time per iteration).

\section{Diffusion Monte Carlo}

\subsection{Basic idea}

While the VMC method is limited by the use of an approximate wave function $\Psi$, the idea of the DMC method~\cite{GriSto-JCP-71,And-JCP-75,And-JCP-76,ReyCepAldLes-JCP-82,MosSchLeeKal-JCP-82} is to sample from the exact wave function $\Psi_0$ of the ground state of the system. If we have this exact wave function $\Psi_0$, then the associated exact energy $E_0$ can be obtained from the mixed expectation value using the trial wave function $\Psi$,
\begin{equation}
E_0 = \frac{\bra{\Psi_0} \hat{H} \ket{\Psi}}{\braket{\Psi_0}{\Psi}} = \frac{\int \d\b{R} \, \Psi_0(\b{R}) \Psi(\b{R}) E_\L(\b{R})}{\int \d\b{R} \, \Psi_0(\b{R}) \Psi(\b{R})},
\label{E0proj}
\end{equation}
since $\Psi_0$ is an eigenfunction of the Hamiltonian $\hat{H}$. The advantage of the mixed expectation value~(\ref{E0proj}) is that it does not require calculating the action of $\hat{H}$ on $\Psi_0$. The integral in Eq.~(\ref{E0proj}) involves the local energy of the trial wave function, $E_\L(\b{R}) = (H\Psi(\b{R}))/\Psi(\b{R})$, and can be estimated in a similar way as in VMC by calculating the average of $E_\L(\b{R})$ on a sample of points $\b{R}_k$ representing the mixed distribution $\Psi_0(\b{R}) \Psi(\b{R})/\int \d\b{R} \, \Psi_0(\b{R}) \Psi(\b{R})$.

But how to access to the exact wave function $\Psi_0$? Let us consider the action of the \textit{imaginary-time evolution operator} ($t \to -i t$) on an arbitrary wave function such as the trial wave function $\Psi$
\begin{equation}
\ket{\Psi(t)} = e^{-(\hat{H} -E_\T)t} \ket{\Psi},
\label{Psit}
\end{equation}
where $E_\T$ is for now an undetermined trial energy. Using the spectral decomposition of the evolution operator (written with the eigenstates $\Psi_i$ and the eigenenergies $E_i$ of $\hat{H}$), we see that the limit of an infinite propagation time is dominated by the state $\Psi_0$ with the lowest energy having a nonzero overlap with $\Psi$
\begin{equation}
\lim_{t\to\infty} \ket{\Psi(t)} = \lim_{t\to\infty} \sum_i e^{-(E_i -E_\T)t} \ket{\Psi_i} \braket{\Psi_i}{\Psi} =  \lim_{t\to\infty} e^{-(E_0 -E_\T)t} \ket{\Psi_0} \braket{\Psi_0}{\Psi},
\end{equation}
since all the other states of energies $E_i >E_0$ decay exponentially faster. The exponential $e^{-(E_0 -E_\T)t}$ can be eliminated by adjusting $E_\T$ to $E_0$, and we then obtain that $\Psi(t)$ becomes proportional to $\Psi_0$
\begin{equation}
\lim_{t\to\infty} \ket{\Psi(t)} \propto  \ket{\Psi_0}.
\end{equation}
In position representation, Eq.~(\ref{Psit}) is written as
\begin{equation}
\Psi(\b{R}_\f,t) = \int \d\b{R}_\i \, G(\b{R}_\f|\b{R}_\i;t) \Psi(\b{R}_\i),
\end{equation}
where $G(\b{R}_\f|\b{R}_\i;t) = \bra{\b{R}_\f} e^{-(\hat{H} -E_\T)t} \ket{\b{R}_\i}$ is called the \textit{Green function} (or the imaginary-time propagator from $\b{R}_\i$ to $\b{R}_\f$). Multiplying and dividing by $\Psi(\b{R}_\f)$ and $\Psi(\b{R}_\i)$, we obtain the evolution equation of the mixed distribution $f(\b{R},t) = \Psi(\b{R},t) \Psi(\b{R})$
\begin{equation}
f(\b{R}_\f,t) = \int \d\b{R}_\i \, \tilde{G}(\b{R}_\f|\b{R}_\i;t) \, \Psi(\b{R}_\i)^2,
\end{equation}
where $\tilde{G}(\b{R}_\f|\b{R}_\i;t)$ is the {\it importance-sampling} Green function,
\begin{equation}
\tilde{G}(\b{R}_\f|\b{R}_\i;t) = \Psi(\b{R}_\f) \, G(\b{R}_\f|\b{R}_\i;t) \, \frac{1}{\Psi(\b{R}_\i)},
\end{equation}
i.e. $\tilde{G}(\b{R}_\f|\b{R}_\i;t)$ is $G(\b{R}_\f|\b{R}_\i;t)$ similarity transformed by the diagonal matrix that has the values of $\Psi$ along the diagonal.
In the limit of infinite time, the mixed distribution becomes proportional to the target stationary distribution: $f(\b{R}) = \lim_{t\to\infty} f(\b{R},t) \propto \Psi_0(\b{R}) \Psi(\b{R})$.

In practice, an analytical expression of the Green function is known only in the limit of a short propagation time, $\tilde{G}(\b{R}_\f|\b{R}_\i;\tau)$, where $\tau$ is a small time step, and one must thus iterate to obtain the stationary distribution
\begin{equation}
f(\b{R}) = \lim_{M\to\infty} \int \d\b{R}_1 \d\b{R}_2 ... \d\b{R}_{M} \, \tilde{G}(\b{R}|\b{R}_{M};\tau) \tilde{G}(\b{R}_{M}|\b{R}_{M-1};\tau) ... \tilde{G}(\b{R}_2|\b{R}_1;\tau) \, \Psi(\b{R}_1)^2.
\label{fRptinf}
\end{equation}
A short-time approximation to the Green function is obtained by applying the Trotter-Suzuki formula, $e^{-\left(\hat{T}+\hat{V}\right)\tau} = e^{-\hat{V}\tau/2} e^{-\hat{T}\tau} e^{-\hat{V}\tau/2} + O(\tau^3)$, where $\hat{T}$ and $\hat{V}$ are the kinetic and potential energy operators. In position representation, this approximation leads to the following expression
\begin{equation}
G(\b{R}_\f|\b{R}_\i;\tau) \approx \frac{1}{\left(2\pi \tau\right)^{3N/2}} e^{- \frac{\left( \b{R}_\f - \b{R}_\i \right)^2}{2\tau}} e^{- \left( \frac{V(\b{R}_\f)+V(\b{R}_\i)}{2} - E_\T\right) \tau},
\end{equation}
where $V(\b{R})$ is the potential energy. Similarly, assuming for now that the trial wave function is of the same sign in $\b{R}_\i$ and $\b{R}_\f$, i.e. $\Psi(\b{R}_\f)/\Psi(\b{R}_\i)>0$, a short-time approximation to the importance-sampling Green function is~\cite{ReyCepAldLes-JCP-82,Sch-INC-87}
\begin{equation}
\tilde{G}(\b{R}_\f|\b{R}_\i;\tau) \approx \frac{1}{\left(2\pi \tau\right)^{3N/2}} e^{- \frac{\left( \b{R}_\f - \b{R}_\i -\b{v}(\b{R}_\i) \tau\right)^2}{2\tau}} e^{- \left( \frac{E_\L(\b{R}_\f)+E_\L(\b{R}_\i)}{2} - E_\T\right) \tau},
\label{Gtshorttime}
\end{equation}
where the drift velocity $\b{v}(\b{R}) = \bm{\nabla} \Psi(\b{R}) / \Psi(\b{R})$ and the local energy $E_\L(\b{R})$ were assumed constant between $\b{R}_\i$ and $\b{R}_\f$. This short-time approximation implies a \textit{finite time-step error} in the calculation of all observables, which should in principle be eliminated by extrapolating the results to $\tau=0$ (see Refs.~\cite{RotVrb-JCP-87,AndGar-JCP-87,ReyOweLes-JCP-87} for proofs that the time-step error vanishes in the $\tau \to 0$ limit).

\subsection{Stochastic realization}

The stochastic realization of Eq.~(\ref{fRptinf}) is less trivial than for VMC. The Green function $\tilde{G}(\b{R}_\f|\b{R}_\i;\tau)$ is generally not a stochastic matrix, since it does not conserve the normalization of the probability density: $\int \d\b{R}_\f \; \tilde{G}(\b{R}_\f|\b{R}_\i;\tau) \not = 1$.
We can nevertheless write the elements of $\tilde{G}$ as the product of the corresponding elements of a stochastic matrix $P$ and a weight matrix $W$,
\begin{equation}
\tilde{G}(\b{R}_\f|\b{R}_\i;\tau) = P(\b{R}_\f|\b{R}_\i) W(\b{R}_\f|\b{R}_\i),
\end{equation}
where, in the short-time approximation, $P(\b{R}_\f|\b{R}_\i) = \left(2\pi \tau\right)^{-3N/2} e^{- \left( \b{R}_\f - \b{R}_\i -\b{v}(\b{R}_\i) \tau\right)^2/2\tau}$ and\\ $W(\b{R}_\f|\b{R}_\i)=e^{- \left( (E_\L(\b{R}_\f)+E_\L(\b{R}_\i))/2 - E_\T\right) \tau}$.
Note that $\tilde{G}$ reduces to a stochastic matrix in the limit $\Psi \to \Psi_0$.
The stochastic realization is then a weighted random walk. Start from a walker at an initial position $\b{R}_1$ with a weight $w_1=1$, i.e., a distribution $w_1 \delta(\b{R}-\b{R}_1)$. Sample the position $\b{R}_2$ of the walker at the next iteration from the probability distribution $P(\b{R}_2|\b{R}_{1})$ [according to Eq.~(\ref{Ri_to_Rf})] and give it weight $w_2 = W(\b{R}_2|\b{R}_{1}) \times w_1$, sample the third position $\b{R}_3$ from the probability distribution $P(\b{R}_3|\b{R}_{2})$ and give it weight $w_3 = W(\b{R}_3|\b{R}_{2}) \times w_2$, and so on. After an equilibration phase, the random walk should sample the stationary distribution $f(\b{R}) \propto \E[w_k \delta(\b{R}-\b{R}_k)] \approx (1/M) \sum_{k=1}^M w_k \delta(\b{R}-\b{R}_k)$. In reality, this procedure is terribly inefficient. Because the weights $w_k$ are products of a large number of random variables, they can become very large at some iterations and very small at other iterations. Consequently, the averages are dominated by a few points with large weights, even though the calculation of any point of the Markov chain takes the same computational time regardless of its weight. This problem can be alleviated by keeping the product of the weights for only a finite number $n$ of consecutive iterations~\cite{CafCla-JCP-88}
\begin{equation}
w_k =  \prod_{l={k-n+1}}^k  W({\bf R}_l|{\bf R}_{l-1}).
\end{equation}
However, using a finite $n$ introduces a bias in the sampled stationary distribution. In practice, for an $n$ large enough to have a reasonably small bias, this procedure still remains inefficient.

The solution is to use at each iteration $k$ a population of $M_k$ walkers, with positions $\b{R}_{k,\alpha}$ and weights $w_{k,\alpha}$ (where $\alpha=1,2,...,M_k$), performing random walks with a {\it branching or birth-death process} designed to make the weights $w_{k,\alpha}$ vary in only a small range from walker to walker in a given iteration, and from iteration to iteration, while still sampling the correct distribution $f(\b{R}) \propto \E[\sum_{\alpha=1}^{M_k} w_{k,\alpha} \delta(\b{R}-\b{R}_{k,\alpha})] \approx (1/M) \sum_{k=1}^M \sum_{\alpha=1}^{M_k} w_{k,\alpha} \delta(\b{R}-\b{R}_{k,\alpha})$. Various unbiased variants are possible, characterized by a population size $M_k$ that either varies or is constant from iteration to iteration, and by weights $w_{k,\alpha}$ that can either be equal or different for each walker.

The simplest variant uses a varying population size $M_k$ and weights all equal to one, $w_{k,\alpha}=1$. At each iteration $k$, each walker $\alpha$ is replaced by $m_{k,\alpha}$ unit-weight copies of itself, where $m_{k,\alpha}$ is an integer equal on average to what should be the current weight $W_{k,\alpha}=W(\b{R}_{k,\alpha}|\b{R}_{k-1,\alpha})$. For example, if the walker $\alpha$ should have the weight $W_{k,\alpha}=2.7$ at iteration $k$, this walker is replaced by $m_{k,\alpha}=3$ copies of itself with a probability 0.7 or replaced by $m_{k,\alpha}=2$ copies of itself with a probability 0.3.
More generally, $m_{k,\alpha}=\lfloor W_{k,\alpha} \rfloor + 1$ with probability $W_{k,\alpha}-\lfloor W_{k,\alpha} \rfloor$ and $m_{k,\alpha}=\lfloor W_{k,\alpha} \rfloor$ otherwise, where $\lfloor W_{k,\alpha} \rfloor$ is the nearest integer smaller than $W_{k,\alpha}$.
If $m_{k,\alpha}=0$ the walker is terminated.  This procedure does not change the sampled stationary distribution
\footnote{One can write: $\E\left[ \sum_{\alpha=1}^{M_k} W_{k,\alpha} \delta (\b{R} -\b{R}_{k,\alpha}) \right] = \E\left[ \sum_{\alpha=1}^{M_k} m_{k,\alpha} \delta (\b{R} -\b{R}_{k,\alpha}) \right] = \E\left[ \sum_{\alpha=1}^{M_{k+1}} \delta (\b{R} -\b{R}_{k+1,\alpha}) \right]$,
where $\b{R}_{k+1,\alpha}$ are the positions of the $M_{k+1}=\sum_{\alpha=1}^{M_k} m_{k,\alpha}$ walkers used for the next iteration $k+1$ obtained after making $m_{k,\alpha}$ copies of the $\alpha^{th}$ walker.}.
This variant has the disadvantage that the integerization of the weights results in unnecessary duplications of walkers, leading to more correlated walkers and thus to a smaller number of statistically independent points in the sample.
Another disadvantage is that it leads to unnecessary fluctuations in the sum of the weights, a quantity that is relevant for computing the growth estimator of the energy.

A better solution is the split-join algorithm~\cite{UmrNigRun-JCP-93} which limits the duplication of walkers by keeping residual noninteger weights $w_{k,\alpha}$. At each iteration $k$, after updating the weights according to $w_{k,\alpha}= W(\b{R}_{k,\alpha}|\b{R}_{k-1,\alpha}) \times w_{k-1,\alpha}$, each walker $\alpha$ with a weight $w_{k,\alpha} > 2$ is split into $\lfloor w_{k,\alpha} \rfloor$ walkers, each being attributed the weight $w_{k,\alpha} / \lfloor w_{k,\alpha} \rfloor$.
If walkers $\alpha$ and $\beta$ each have weight $<1/2$, keep walker $\alpha$ with probability $w_{k,\alpha}/(w_{k,\alpha} + w_{\beta,k})$ and walker $\beta$ otherwise. In either case, the surviving walker gets weight, $w_{k,\alpha} + w_{\beta,k}$. This algorithm has the advantage that it conserves the total weight of the population of walkers $W_k = \sum_{\alpha=1}^{M_k} w_{k,\alpha}$ for a given iteration.
Yet another possibility is the stochastic reconfiguration algorithm~\cite{CalSor-PRB-98,AssCafKhe-PRE-00}, which uses a fixed population size $M_k$, and walkers of equal noninteger weights within each iteration, though the weights of the walkers fluctuate from one iteration to the next.

To avoid the explosion or extinction of the population of walkers (or their weights if $M_k$ is kept fixed), the value of $E_\T$ can be adjusted during the iterations. For example, a choice for $E_\T$ at iteration $k+1$ is $E_\T(k+1) = E_0^{\text{est}}(k) - C \log (W_k/W_0)$ where $E_0^{\text{est}}(k)$ is  an estimate of $E_0$ at iteration $k$, $C$ is a constant, $W_k$ is the total weight of the population of walkers and $W_0$ is the target total weight. Because of fluctuations, $E_\T$ thus slightly varies with respect to $E_0$ during the iterations, which introduces a systematic bias on the weights and thus on the stationary distribution $f(\b{R})$. The adjustment of $E_\T$ makes $f(\b{R})$ too small in regions where $E_\L(\b{R})<E_0$ and too large in regions where $E_\L(\b{R})>E_0$. Both of these have the effect of raising the energy.  This is called \textit{population-control error}. This error is generally small and decreases with increasing number of walkers as $1/M_k$~\cite{UmrNigRun-JCP-93}.  Besides, it is possible to eliminate almost completely this error by undoing the modification of weights introduced by the variation of $E_\T$ for the last several iterations~\cite{NigBlo-PRB-86,UmrNigRun-JCP-93}.

In the limit of an infinitesimal time step, the transition matrix $P(\b{R}_\f|\b{R}_\i)$ has a stationary distribution $\Psi(\b{R})^2$, and the weight term $W(\b{R}_\f|\b{R}_\i)$ converts this distribution into the mixed distribution $\Psi_0(\b{R}) \Psi(\b{R})$. One can get rid of the finite time-step error in the transition matrix $P(\b{R}_\f|\b{R}_\i)$ by introducing an accept/reject step as in the Metropolis-Hastings algorithm~\cite{ReyCepAldLes-JCP-82}. For this, the transition matrix is redefined as $P(\b{R}_\f|\b{R}_\i) = P_{\text{acc}}(\b{R}_\f|\b{R}_\i) P_{\text{prop}}(\b{R}_\f|\b{R}_\i)$, for $\b{R}_\i \not = \b{R}_\f$, with the proposal probability
\begin{equation}
P_{\text{prop}}(\b{R}_\f|\b{R}_{\i}) = \frac{1}{\left(2\pi \tau\right)^{3N/2}} e^{- \frac{\left( \b{R}_\f - \b{R}_\i -\b{v}(\b{R}_\i) \tau\right)^2}{2\tau}},
\label{PpropDMC}
\end{equation}
and the acceptance probability
\begin{equation}
P_{\text{acc}}(\b{R}_\f|\b{R}_{\i}) = \min \left\{ 1, \frac{P_{\text{prop}}(\b{R}_\i|\b{R}_{\f}) \Psi(\b{R}_{\f})^2}{P_{\text{prop}}(\b{R}_\f|\b{R}_{\i}) \Psi(\b{R}_{\i})^2} \right\}.
\end{equation}
With this modification, $P(\b{R}_\f|\b{R}_\i)$ has the stationary distribution $\Psi(\b{R})^2$ even for a finite time step. Of course, the finite time-step error persists in the term $W(\b{R}_\f|\b{R}_\i)$. Since certain moves are rejected, $P(\b{R}_\f|\b{R}_\i)$ corresponds now to a process of diffusion with drift with an effective time step $\tau_{\text{eff}}<\tau$. This effective time step can be estimated during the calculation from the average acceptance rate and it is consistent to use it in the term $W(\b{R}_\f|\b{R}_\i)$ in place of $\tau$. In practice, just as in VMC, it is also more efficient in DMC to move the electrons one by one, i.e. to decompose $P(\b{R}_\f|\b{R}_\i)$ according to Eq.~(\ref{Ponebyone}). We then arrive at a DMC algorithm very similar to the VMC algorithm, with weights in addition. Note, however, that since a relatively small time step must be used in DMC, the average moves are smaller than in VMC and the autocorrelation time of the local energy is larger than in VMC.

The energy is calculated as the average of the local energy over the distribution $f(\b{R})/\int \d\b{R} f(\b{R})$. For $M$ iterations (after the equilibration phase) and $M_k$ walkers, we have
\begin{equation}
E_0 \approx \overline{E}_\L = \frac{\sum_{k=1}^{M} \sum_{\alpha=1}^{M_k} w_{k,\alpha} E_\L(\b{R}_{k,\alpha})}{\sum_{k=1}^{M} \sum_{\alpha=1}^{M_k} w_{k,\alpha}}.
\label{E0DMCweights}
\end{equation}
Just as in VMC, there is a zero-variance principle on the energy in DMC. In the limit that the trial wave function $\Psi$ is an exact eigenfunction of the Hamiltonian, $E_\L$ is independent of $\b{R}$, the weights reduce to $1$, and the variance on $\overline{E}_\L$ vanishes.

Note that for an observable $\hat{O}$ that does not commute with the Hamiltonian, the average $\overline{O}_\L$ over the mixed DMC distribution is an estimator of $\bra{\Psi_0}\hat{O}\ket{\Psi}/\braket{\Psi_0}{\Psi}$ which is only an approximation to the exact expectation value $\bra{\Psi_0}\hat{O}\ket{\Psi_0}/\braket{\Psi_0}{\Psi_0}$ with an ${\cal O} (||\Psi-\Psi_0||)$ error. Since the average $\overline{O}_\L$ over the VMC distribution also has an error that is linear in $||\Psi-\Psi_0||$ but with a prefactor that is twice as large, an ${\cal O} (||\Psi-\Psi_0||^2)$ approximation is provided by twice the average of $O_\L$ over the mixed DMC distribution minus the average of $O_\L$ over the VMC distribution~\cite{CepKal-INC-79}. For a recent survey of exact methods for sampling the pure distribution $\Psi_0^2$, see Ref.~\cite{Rot-CJC-13}, and for a discussion of the techniques used for calculating pure expectation values of various classes of operators see Ref.~\cite{NigUmr-INC-98}.

\subsection{Fermionic sign problem}

In Eq.~(\ref{Gtshorttime}), we have assumed that the trial wave function $\Psi(\b{R})$ is always of the same sign, i.e. that it does not have any nodes (points $\b{R}$ so that $\Psi(\b{R})=0$). This is the case for the ground-state wave function of a Bosonic system, and for a few simple electronic systems (two electrons in a spin-singlet state, such as the ground state of the He atom or of the H$_2$ molecule). In this case, the algorithm presented above allows one to obtain the exact energy of the system, after elimination of the finite time-step error and the population-control error.
If the wave function of the Fermionic ground state has nodes, then there is always at least one Bosonic state of lower energy, and the true ground state of the Hamiltonian is a Bosonic state for which the wave function $\Psi_\text{B}(\b{R})$ can be chosen strictly positive. If one applied the Green function exactly, starting from the distribution $\Psi(\b{R})^2$ the distribution would correctly converge to $\Psi_0(\b{R})\Psi(\b{R})$ since the trial wave function is antisymmetric (with respect to the exchange of two electrons) and has a zero overlap with all the Bosonic states which are symmetric. However, in reality one applies the Green function using a finite sampling in position space which does not allow one to impose the antisymmetry. Starting from an antisymmetric wave function $\Psi$, a small component of $\Psi_\text{B}$ can thus appear, and it grows and eventually dominates. The distribution tends to $\Psi_\text{B}(\b{R})\Psi(\b{R})$ and the energy formula in Eq.~(\ref{E0DMCweights}) only gives $0/0$ (the positive and negative weights cancel out) with statistical noise. Even if one imposed antisymmetry and eliminated the Bosonic states, e.g. by considering all electron permutations in each walker, the problem persists because different paths between the same points in this antisymmetrized space can contribute with opposite sign. Since $\Psi_0$ and $-\Psi_0$ are equally good solutions of the Schr\"odinger equation, the algorithm would sample each with approximately equal probability, leading again to the cancellation of positive and negative weight contributions. These are different manifestations of the infamous \textit{Fermionic sign problem}.

\subsection{Fixed-node approximation}

To avoid the sign problem in DMC, the \textit{fixed-node approximation} (FN)~\cite{And-JCP-75,KlePic-JCP-76,And-JCP-76} is introduced. The idea is to force the convergence to a wave function approximating the Fermionic ground state by fixing its nodes to be the same as those of the trial wave function $\Psi(\b{R})$. Formally, one can define the FN Hamiltonian, $\hat{H}_{\FN}$, by adding to the true Hamiltonian $\hat{H}$ infinite potential barriers at the location of the nodes of $\Psi(\b{R})$~\cite{BadHayNee-PRB-08}. The ground-state wave function of this Hamiltonian is called the FN wave function $\Psi_{\FN}$ and its energy is the FN energy $E_{\FN}$,
\begin{equation}
\hat{H}_{\FN} \ket{\Psi_{\FN}} = E_{\FN} \ket{\Psi_{\FN}}.
\end{equation}
In the $3N$-dimensional space of positions $\b{R}$, the nodes of $\Psi(\b{R})$ define hypersurfaces of dimension $3N-1$. The position space is then partitioned in nodal pockets of $\Psi(\b{R})$, delimited by nodal surfaces, in which the wave function has a fixed sign. In each nodal pocket, the FN wave function is the solution to the Schr\"odinger equation satisfying vanishing boundary conditions on the nodal surface. The FN Green function corresponding to the Hamiltonian $\hat{H}_\FN$ is
\begin{equation}
G_{\FN}(\b{R}_\f|\b{R}_\i;t) = \bra{\b{R}_\f} e^{-(\hat{H}_{\FN} -E_\T)t} \ket{\b{R}_\i},
\end{equation}
and only permits moves $\b{R}_\i \to \b{R}_\f$ inside a nodal pocket. The importance-sampling FN Green function,
\begin{equation}
\tilde{G}_{\FN}(\b{R}_\f|\b{R}_\i;t)=\Psi(\b{R}_\f) \; G_{\FN}(\b{R}_\f|\b{R}_\i;t) \; \frac{1}{\Psi(\b{R}_\i)},
\end{equation}
also confines the moves inside a nodal pocket, and it is thus always greater or equal to zero. A short-time approximation to $\tilde{G}_{\FN}(\b{R}_\f|\b{R}_\i;t)$ is then again given by Eq.~(\ref{Gtshorttime}). The stochastic algorithm previously described can thus be applied directly. Thanks to the FN approximation, the weights always remain positive, and the stationary mixed distribution $f(\b{R})$ is proportional to $\Psi_{\FN}(\b{R}) \Psi(\b{R})$.

The largest contributions to the finite time-step error come from singularities of the drift velocity $\b{v}(\b{R})=\bm{\nabla} \Psi(\b{R})/\Psi(\b{R})$ and of the local energy $E_\L(\b{R})$ in the Green function of Eq.~(\ref{Gtshorttime}). Since the gradient of the trial wave function $\bm{\nabla} \Psi(\b{R})$ (and of the exact wave function) does not generally vanish at the location of the nodes, the drift velocity $\b{v}(\b{R})$ diverges at the nodes, which leads to too large moves near the nodes for finite time steps. The drift velocity has discontinuities also at particle coalescences (both electron-nucleus and electron-electron). Similarly, for an approximate trial wave function $\Psi(\b{R})$, the local energy $E_\L(\b{R})$ also diverges at the nodes and at particle coalescences (unless the Kato cusp conditions~\cite{Kat-CPAM-57,PacBye-JCP-66} are imposed). The finite time-step error can be greatly reduced by replacing $\b{v}(\b{R})$ and $E_\L(\b{R})$ in the Green function by approximate integrals of these quantities over the time step $\tau$~\cite{UmrNigRun-JCP-93}.

If importance sampling is not used, it is necessary to kill walkers that cross the nodes of $\Psi$ to impose the FN boundary condition.
In practice importance sampling is almost always used.
In that case, it is better to reject the moves of walkers crossing the nodes, which is consistent with the FN approximation, but even this is not necessary
since the number of walkers that cross the node per unit time goes to zero as $\tau \to 0$~\cite{UmrNigRun-JCP-93}\footnote{
The drift velocity moves electrons away from the nodal surface, but for small $\tau$ the diffusion term dominates and can cause walkers to cross nodes.
The density of walkers goes quadratically to zero near nodes and walkers that are roughly within a distance $\sqrt{\tau}$ can cross.
Hence the number that cross per Monte Carlo step goes as $\int_0^{\sqrt{\tau}} x^2 \d x \sim \tau^{3/2}$, and so the number that cross
per unit time goes to zero as $\sqrt{\tau}$.}.
For a finite time step, there are node crossing events, but these are just part of the finite time-step error and in practice essentially the same time-step error is obtained whether the walkers are allowed to cross nodes or not.

We may wonder whether the walkers have to sample all the nodal pockets. The tiling theorem~\cite{Cep-JSP-91} establishes that all the nodal pockets of the ground-state wave function of a many-electron Hamiltonian with a reasonable local potential are equivalent, i.e., the permutations of any nodal pocket are sufficient to cover the entire space. This means that, for ground-state calculations, the distribution of the walkers over the nodal pockets is irrelevant.

By applying the variational principle, it is easy to show that the FN energy is an upper bound to the exact energy
\begin{equation}
E_{\FN} = \frac{\bra{\Psi_\FN} \hat{H}_{\FN} \ket{\Psi_\FN}}{\braket{\Psi_\FN}{\Psi_\FN}} = \frac{\bra{\Psi_\FN} \hat{H} \ket{\Psi_\FN}}{\braket{\Psi_\FN}{\Psi_\FN}} \geq E_0,
\end{equation}
the second equality coming from the fact that the infinite potential barriers in $\hat{H}_{\FN}$ do not contribute to the expectation value since $\Psi_\FN$ is zero on the nodal surface. Since the wave function $\Psi_\FN$ is an eigenfunction of $\hat{H}_{\FN}$, the FN energy can also be expressed using the mixed expectation value
\begin{equation}
E_{\FN} = \frac{\bra{\Psi_\FN} \hat{H}_{\FN} \ket{\Psi}}{\braket{\Psi_\FN}{\Psi}} = \frac{\bra{\Psi_\FN} \hat{H} \ket{\Psi}}{\braket{\Psi_\FN}{\Psi}},
\label{EFNproj}
\end{equation}
where the Hamiltonian $\hat{H}_{\FN}$ has been replaced by $\hat{H}$ for essentially the same reason as before, viz., both $\Psi$ and $\Psi_\FN$ are zero where $\hat{H}_{\FN}$ is infinite. In practice, the FN energy is thus obtained by the same energy formula~(\ref{E0DMCweights}).

The accuracy of the DMC results with the FN approximation thus depends on the quality of the nodal surface of the trial wave function. For a trial wave function with a single Slater determinant, the error due to the FN approximation can often be large, even for energy differences.
For example, for the C$_2$ molecule, the FN error for a single-determinant trial wave function is 1.6 eV for the total energy and 0.8 eV for the dissociation energy~\cite{TouUmr-JCP-08}.
In order to reduce this error, one can use several Slater determinants and optimize  the parameters of the wave function (Jastrow parameters, coefficients of determinants, coefficients that express the orbitals in terms of the basis functions, and exponents of the basis functions) in VMC (see Refs.~\cite{UmrWilWil-PRL-88,NigMel-PRL-01,SchFil-JCP-04,UmrFil-PRL-05,Sor-PRB-05,UmrTouFilSorHen-PRL-07,TouUmr-JCP-07,TouUmr-JCP-08}). This allows one to reach near chemical accuracy ($\sim$ 1 kcal/mol) in DMC for calculations of energy differences such as molecular atomization energies~\cite{PetTouUmr-JCP-12}.

\newpage
\appendix
\section*{Appendix: Statistical estimator of nonlinear functions of expectation values}
\addcontentsline{toc}{section}{Appendix: Statistical estimator of nonlinear functions of expectation values}

We often need to estimate nonlinear functions of expectation values. The simplest example is the variance,
\begin{equation}
\V[X] = \E[X^2]-\E[X]^2,
\end{equation}
which is a quadratic function of the expectation values of two random variables $X^2$ and $X$. Another example is the calculation of the energy in DMC using weights [see Eq.~(\ref{E0DMCweights})], with simplified notation,
\begin{equation}
E_0 = \frac{\E[w E_\L]}{\E[w]},
\end{equation}
involving a ratio of two expectation values.

Consider a nonlinear function, $f(\E[X],\E[Y])$, of two expectation values, $\E[X]$ and $\E[Y]$.
The usual simple estimator of $f(\E[X],\E[Y])$ is $f(\overline{X},\overline{Y})$ where
\begin{equation}
\overline{X} = \frac{1}{M_{\block}} \sum_{b=1}^{M_\block} \overline{X}_b,
\end{equation}
and
\begin{equation}
\overline{Y} = \frac{1}{M_{\block}} \sum_{b=1}^{M_\block} \overline{Y}_b,
\end{equation}
are averages over a finite number of blocks $M_\block$, and $\overline{X}_b$ and $\overline{Y}_b$ are the block averages of $X$ and $Y$, respectively [see Eq.~(\ref{Xblock})].
As discussed before, each block average is itself an average over a sufficiently large number of steps, $M_\s$, so that the block averages can be assumed to be independent of each other.
The simple estimator can be justified as follows. (i) When the law of large numbers holds, $\overline{X}$ and $\overline{Y}$ converge, with increasing $M_\block$, almost surely to $\E[X]$ and $\E[Y]$, respectively. (ii) Hence, $f(\overline{X},\overline{Y})$ converges to $f(\E[X],\E[Y])$ provided that $f$ is continuous at the point $(\E[X],\E[Y])$.
However, because $f$ is nonlinear, $f(\overline{X},\overline{Y})$ has a systematic error, i.e. $E[f(\overline{X},\overline{Y})] \ne f(\E[X],\E[Y])$, that vanishes only in the limit of infinite sample size, $M_\block \to \infty$. 
Though not necessary, in the following, for the sake of simplicity, we assume that $f(\overline{X},\overline{Y})$ has a finite expectation value and a finite variance\footnote{
$E[f(\overline{X},\overline{Y})]$ can be undefined when $f$ has a point at which it diverges, e.g.,
$f(x,y)=x/y$.  In this case, this definition of the systematic error does not have a strict meaning.
In practice, this is not a problem for this $f$ provided that the absolute value of the expectation value
of $Y$ over a block is larger than a few times the square root of its variance, say,
$|E[\overline{Y_\block}]| > 5 \sqrt{V[\overline{Y_\block}]}$.
}.

\subsection*{Systematic error}

Let us first consider the case of a nonlinear function
$f(x)$ of a single variable. By definition, the systematic error of the estimator $f(\overline{X})$ is $\E[f(\overline{X})] - f(\E[X])$. The systematic error can be evaluated using a second-order Taylor expansion of the function $f(\overline{X})$ around $\E[X]$ (assuming that $f$ is at least a $C^2$ function in the neighborhood of $\E[X]$)
\begin{equation}
f(\overline{X}) = f(\E[X]) + \left( \frac{\d f}{\d x} \right)\; (\overline{X}-\E[X]) + \frac{1}{2} \left(\frac{\d^2 f}{\d x^2} \right)\; (\overline{X}-\E[X])^2 + \cdots,
\label{fX}
\end{equation}
where the derivatives of $f$ are evaluated at $\E[X]$.
If we take the expectation value of this expression, the linear term vanishes
\begin{equation}
\E[f(\overline{X})] = f(\E[X]) + \frac{1}{2} \left(\frac{\d^2 f}{\d x^2} \right)\; \E\!\left[(\overline{X}-\E[X])^2\right] + \cdots.
\label{EfX}
\end{equation}
Assuming the random variable $X$ has a finite variance and that the higher-order terms can be neglected, the systematic error is thus
\begin{equation}
\E[f(\overline{X})] - f(\E[X]) \;=\; \frac{1}{2} \left(\frac{\d^2 f}{\d x^2} \right)\; \V[\overline{X}] + \cdots \;=\; \frac{1}{2} \left(\frac{\d^2 f}{\d x^2} \right)\; \frac{\V[\overline{X}_\block]}{M_\block} + \cdots.
\label{EfXV}
\end{equation}
Hence, the estimator $f(\overline{X})$ has a systematic error with a leading term proportional to $1/M_{\block}$. Note that if the hypotheses (especially the finite variance) are not satisfied, the systematic error can decrease more slowly than $1/M_\block$. Equation~(\ref{EfXV}) can easily be generalized to a function of several variables. For example, for two variables, the systematic error is
\begin{eqnarray}
\E[f(\overline{X},\overline{Y})] - f(\E[X],\E[Y]) &=& \frac{1}{2} \left(\frac{\partial^2 f}{\partial x^2} \right)\; \frac{\V[\overline{X}_\block]}{M_\block} + \frac{1}{2} \left(\frac{\partial^2 f}{\partial y^2} \right)\; \frac{\V[\overline{Y}_\block]}{M_\block}
\nonumber\\
&&+  \left(\frac{\partial^2 f}{\partial x \partial y} \right) \; \frac{\Cov[\overline{X}_\block,\overline{Y}_\block]}{M_\block} + \cdots,
\label{EfXY}
\end{eqnarray}
where the second-order derivatives are evaluated at $(\E[X],\E[Y])$.
The leading neglected term is $O(1/M_\block^2)$ if the third moments of $X$ and $Y$ are finite. The second-order derivatives in Eq.~(\ref{EfXY}) can in practice be evaluated at $(\overline{X}, \overline{Y})$ without changing the order of the approximation. Hence, an estimator for $f(\E[X],\E[Y])$ with an $O(1/M_\block^2)$ error is
\begin{eqnarray}
f(\E[X],\E[Y]) &\approx& f(\overline{X},\overline{Y}) - \frac{1}{2} \left(\frac{\partial^2 f}{\partial x^2} \right)\; \frac{\V[\overline{X}_\block]}{M_\block} - \frac{1}{2} \left(\frac{\partial^2 f}{\partial y^2} \right)\; \frac{\V[\overline{Y}_\block]}{M_\block}
\nonumber\\
&&-  \left(\frac{\partial^2 f}{\partial x \partial y} \right) \; \frac{\Cov[\overline{X}_\block,\overline{Y}_\block]}{M_\block} + \cdots,
\label{fXYfinal}
\end{eqnarray}
where the second-order derivatives are evaluated at $(\overline{X}, \overline{Y})$.

This approach is general and can be used to recover some well-known unbiased estimators. For example, let us consider the covariance of two random variables
\begin{eqnarray}
 \Cov[X,Y] \;=\; \E[X Y]-\E[X] \E[Y] \;=\; f(\E[X Y],\E[X],\E[Y]),
\end{eqnarray}
for which $f(x,y,z)=x-yz$. In this case, the generalization of Eq.~(\ref{EfXY}) to three variables with $\overline{X} = (1/M) \sum_{i=1}^M X_i$ and $\overline{Y} = (1/M) \sum_{i=1}^M Y_i$ where $X_i$ and $Y_i$ are $M$ uncorrelated realizations of $X$ and $Y$, respectively, gives
\begin{equation}
\E[\overline{X Y}-\overline{X} \; \overline{Y}] - \Cov[X,Y] \;=\; - \frac{\Cov[X,Y]}{M},
\label{syst2p}
\end{equation}
which leads to the usual unbiased estimator for the covariance
\begin{equation}
\Cov[X,Y] \;\approx\; \frac{M}{M -1} \left( \overline{X Y}-\overline{X} \; \overline{Y} \right)
\;=\; \frac{1}{M-1} \sum_{i=1}^{M} \left( X_i - \overline{X} \right) \left( Y_i - \overline{Y} \right).
\label{cov}
\end{equation}

\subsection*{Statistical uncertainty}

First consider a function of a single variable. The statistical uncertainty of $f(\overline{X})$ is given by $\sigma[f(\overline{X})]=\sqrt{\V[f(\overline{X})]}$ where the variance of $f(\overline{X})$ is $\V[f(\overline{X})] =  \E\left[ \left(f(\overline{X}) - \E[f(\overline{X})] \right)^2\right]$. Subtracting Eq.~(\ref{EfX}) from Eq.~(\ref{fX}) gives
\begin{equation}
f(\overline{X}) - \E[f(\overline{X})] = \left(\frac{\d f}{\d x} \right)\; (\overline{X}-\E[X]) + \cdots.
\end{equation}
Taking the square of this equation and the expectation value leads to the leading term in the variance of $f(\overline{X})$
\begin{equation}
\V[f(\overline{X})] = \left(\frac{\d f}{\d x} \right)^2 \; \V[\overline{X}] + \cdots.
\end{equation}
This equation can be generalized to a function of several variables.
For example, for two variables, the variance of $f(\overline{X},\overline{Y})$ is
\begin{equation}
\V[f(\overline{X},\overline{Y})] = \left(\frac{\partial f}{\partial x} \right)^2 \; \V[\overline{X}] + \left(\frac{\partial f}{\partial y} \right)^2 \; \V[\overline{Y}] + 2 \left(\frac{\partial f}{\partial x} \right)  \left(\frac{\partial f}{\partial y} \right) \Cov[\overline{X},\overline{Y}] +\cdots.
\label{VfXY}
\end{equation}
Equation~(\ref{VfXY}) can be used for estimating the variance of $f(\overline{X},\overline{Y})$ at the cost of evaluating the variances $\V[\overline{X}]$ and $\V[\overline{Y}]$ and the covariance $\Cov[\overline{X},\overline{Y}]$.
Note however, that it can give a severe underestimate of the error if ${\partial f}/{\partial x}$
and ${\partial f}/{\partial y}$ are small and $M_\block$ is not sufficiently large.

There is a simple alternative for estimating the variance of $f$ that does not suffer from this limitation
and does not require calculating covariances.
Consider again the case of a single variable.
Instead of defining the  block average of $f$ in the obvious way, i.e. $\overline{f}_b   =  f(\overline{X}_b)$,
we define the block average of $f$ as
\begin{eqnarray}
\overline{f}_1 & = & f(\overline{X}_1) \text{ \ \ for the first block $b=1$}
\nonumber\\
\overline{f}_b  & = & b f(\overline{X}(b)) - (b-1)f (\overline{X}(b-1)) \text{ \ \ for any block } b \geq 2,
\label{lin}
\end{eqnarray}
where $\overline{X}(b)$ is the running global average up to block $b$
\begin{equation}
\overline{X}(b) = \frac{1}{b} \sum_{b'=1}^b \overline{X}_{b'}.
\end{equation}
With this definition of the block average, it is easy to check that
\begin{equation}
f({\overline{X}}) = \frac{1}{M_\block}\sum_{b=1}^{M_\block} \overline{f}_b,
\end{equation}
i.e. we have written $f(\overline{X})$ as an average of random variables $\overline{f}_b$. Provided that the variance of $X$ is finite, the block average $\overline{f_b}$ introduced in Eq.~(\ref{lin}) can be expanded as
\begin{equation}
\overline{f}_b =  f(\E[X])+ \left(\frac{\d f}{\d x} \right)\; (\overline{X}_b -\E[X]) + \cdots.
\label{fbinf}
\end{equation}
Assuming that $f$ has a second-order Taylor expansion, the neglected term converges to zero in probability for large $b$, at least as $1/(b M_\s)$.
Therefore, according to Eq.~(\ref{fbinf}), for large $b$, the random variables $\overline{f}_b$ converge to independent and equidistributed random variables (since the random variables $\overline{X}_b$ are)\footnote{The naive definition of the block average as $\overline{f}_b  = f(\overline{X}_b)$ would also lead to Eq.~(\ref{fbinf}) but the neglected term would not  converge to zero for large $b$.}.
Consequently, the variance of $f(\overline{X})$ can be estimated with the usual formula
\begin{equation}
\V[f(\overline{X})] \approx \frac{\V[\overline{f}_b]}{M_\block} \approx \frac{1}{M_\block-1} \left( \frac{1}{M_\block}\sum_{b=1}^{M_\block} \overline{f}_b^2 - f(\overline{X})^2\right).
\label{VfXsimple}
\end{equation}
This formula applies similarly for functions of several variables. The advantage of Eq.~(\ref{VfXsimple}) over Eq.~(\ref{VfXY})
for estimating the variance is that it is much simpler to implement and compute, especially for functions of many variables. The estimation of the variance can be simply updated at each block, just as for the expectation value.

% BIBLIOGRAPHY---------------------------------------------
\newpage
\renewcommand\refname{References}

\end{document}